\documentclass[a4paper,fleqn,usenatbib]{mnras}

\usepackage[T1]{fontenc}
\usepackage{ae,aecompl}

\usepackage{graphicx}	% Including figure files
\usepackage{amsmath}	% Advanced maths commands
\usepackage{amssymb}	% Extra maths symbols
\usepackage{multicol}        % Multi-column entries in tables
\usepackage{bm}		% Bold maths symbols, including upright Greek
\usepackage{color}
\usepackage{xcolor}
\usepackage{soul}         

\title[Classification of {\em TESS} variable stars]
{Identification and classification of {\em TESS} variable stars}
\author[L.A. Balona]{L. A. Balona\thanks{E-mail: lab@saao.ac.za}\\
South African Astronomical Observatory, P.O. Box 9, Observatory 7935, Cape
Town, South Africa}

\begin{document}

\date{Accepted .... Received ...}

\pagerange{\pageref{firstpage}--\pageref{lastpage}} \pubyear{2011}

\maketitle

\label{firstpage}

\begin{abstract}
Visual classification of the variability classes of over 120\,000 {\em Kepler, K2} 
and {\em TESS} stars is presented.  The sample is mainly based on stars with 
known spectral types. Since variability classification often requires the
location of the star in the H--R diagram, a catalog of effective temperatures 
was compiled. Luminosities were estimated from {\em Gaia DR3}
parallaxes.  The different classes of variable found in this survey are
discussed.  Examples of light curves and periodograms for common variability
classes are shown. A catalogue of projected rotational velocities is also
included.
\end{abstract}


\section{Introduction}

The first step in any scientific investigation of a group of objects with
unknown properties is to look for similarities among them and develop a
classification system. One may then begin to investigate the physical 
circumstances which could lead to the observed properties.  In astrophysics,
the classification of stars into different variability classes has been one
of the most important steps in understanding stellar structure and
evolution.

Until a decade ago, practically all available information on stellar
variability was obtained through ground-based observations and classification
by visual inspection of the light curves.  These results were collected in the 
{\em General Catalogue of Variable Stars} (GCVS, \citealt{Samus2017}).  Over
the years, the definition of the various variability classes has evolved. 
Some classes have fallen away or merged, while new ones have been
discovered.  The increase in data resulting from space photometry has, however,
overwhelmed the capacity of the GCVS.

Given that there are now many millions of stars observed from space with light 
curves of unprecedented accuracy, variability classification has become a 
daunting task.  There is a pressing need for automation, which requires a 
precise definition of each variability class.  The danger is that unexpected 
classes of variability might escape attention.  There is also a need 
for a large sample of stars classified by visual inspection to act as a
learning sample for an artificial intelligence code.  Automated variable
star classification using ground-based photometry was performed by
\citet{Pojmanski2005}, and in the {\em Kepler} field by \citet{Blomme2010, 
Blomme2011} and also by \citet{Audenaert2021} for {\rm TESS} data.

It is evident that a certain amount of familiarity with the appearance of the
light curves and periodograms for different variable star classes is
an important requirement for visual classification.  The author has 
been active in the variable star community for several decades, and has 
visually classified over 20\,000 stars observed by {\em Kepler} and {\em K2}.
The lessons learnt in this exercise were very valuable.

Before embarking on this project, it was necessary to place limitations on the
number of objects to be classified.  The initial goal was to include stars
with known spectral types, since knowledge of the spectral type is often
crucial in allocating a variability class.   A further restriction was made
by limiting the stars in the sample to 12.5\,mag or brighter.  Since the
author's main interest lies in the hot main sequence, the sample was further
restricted to stars earlier than G.  Using the SIMBAD website 
\citep{Wenger2000}, a master catalogue was constructed for stars in the 
{\em Kepler} field.  With the advent of {\em TESS}, this was expanded to an 
all-sky catalogue of 718\,000 stars, which includes many fainter stars and
stars of interest with no spectral type. 

The master catalogue includes the RA, Dec, star names, variability class, 
$UBV$ magnitudes, Str\"{o}mgren $uvby\beta$, effective temperature, luminosity, 
extinction, projected rotation velocity, rotation period, peculiarity code and 
spectral type.  In this paper the master catalogue is not presented, because
only a subset of these stars, i.e. stars observed by {\em Kepler, K2} and 
{\em TESS}, have been classified by the author. 

While the master catalogue contains 718\,000 stars, the number 
of stars actually selected for classification is much smaller: about 20\,000 
stars observed by {\em Kepler} and {\em K2} and about 100\,000 stars observed by 
{\em TESS}.  The main result of this paper is a presentation of variability 
classification of these stars together with the best estimate of their
luminosities and effective temperatures.  The luminosity of each star was 
derived from {\em GAIA DR3} parallaxes \citep{Gaia2016, Gaia2022}.  

Since the variability class depends on the location of the star in the H--R 
diagram, approximate effective temperatures and luminosities are essential.   
For this purpose, and also because the effective temperature  and luminosity
are fundamental parameters in any study, the literature was searched for sources 
of  effective temperature.  The resulting catalogue contains over 900\,000 
entries for nearly 635\,000 stars and is not limited to the {\em Kepler, K2} or
{\em TESS} stars.

Another important parameter is the projected rotational velocity, $v\sin i$.  
The last update of this catalogue \citep{Glebocki2005b} is quite old and an 
updated version with over 75\,000 entries for about 52\,000 stars is presented.
Like the effective temperature catalogue, the rotational velocity catalog
is not restricted to {\em Kepler, K2} or {\em TESS} stars.

\section{Effective temperatures, luminosities and rotation}

There are several ways in which $T_{\rm eff}$ may be estimated and not all
should be treated with equal weight. The most reliable method of obtaining 
$T_{\rm eff}$ is by fitting spectral line profiles using suitable atmospheric 
models.    Spectroscopic modelling may be restricted to the Balmer lines, 
which is less expensive in resources, but may also include other lines.  It is 
clear values of $T_{\rm eff}$ based on spectroscopic fitting should be used in 
preference to any other method.

Narrow-band photometry, such as the Str\"{o}mgren and Geneva systems,
in combination with an appropriate calibration, is a well-known technique for
deriving precise $T_{\rm eff}$ estimates.  Interferometry, when available, can 
also be a very useful source of reliable effective temperatures.  If a
spectroscopic estimate is not available, these estimates of $T_{\rm eff}$,
if available, are to be preferred.

Another method of obtaining $T_{\rm eff}$ is by fitting several photometric
measurements at different wavelengths to a model (the spectral energy
distribution, SED).  While this is a good method, the $U$ band needs to be
included if it is to be used for B stars.  Johnson $UBV$ photometry can be
used to derive  the reddening for B stars using the Q method. The
de-reddened colours then provide a good estimate of $T_{\rm eff}$.

Values of $T_{\rm eff}$ in the {\em Kepler Input Catalogue} 
(KIC, \citealt{Brown2011a}) are derived from 2MASS photometry as well as 
Sloan filters, such as the {\em griz} filters.  The effective temperatures in 
the {\em TESS Input Catalogue} (TIC, \citealt{Stassun2018}) are derived from 
the {\em GAIA DR2} catalog as a base together with a merger of a large number 
of other photometric catalogs, including 2MASS, UCAC4, APASS, SDSS, WISE, etc.  
Care are must be taken in using the KIC and TIC values of $T_{\rm eff}$
for B stars because $U$-band measurements are generally absent.

Finally, if the spectral type and class is known, a crude value of $T_{\rm
eff}$ may be inferred using a table of effective temperatures as a function
of spectral type and spectral class.  This is very often sufficient for
breaking an ambiguity in the variability class.

For each entry in the effective temperature catalogue, the method used to 
derive $T_{\rm eff}$ is recorded by a code. For example, spectroscopic values 
of $T_{\rm eff}$ are given the code {\tt SPC}, estimates from Str\"{o}mgren 
photometry the code {\tt STR}, estimates based on the SED have code {\tt SED}, 
those from the {\em TESS Input Catalogue} have code {\tt TIC}, etc.  Each code 
can be assigned a priority level.  The software calculates the ``best''
value of $T_{\rm eff}$ using the priority level, in accordance with the 
guidelines mentioned above.  For example, if several entries of $T_{\rm eff}$ 
are present for the same star, some with the {\tt SPC} code and others with 
the {\tt SED} and {\tt TIC} codes, it is clear that only those measurements 
with the {\tt SPC} code should be used.  The others should be ignored and the
``best'' value of $T_{\rm eff}$ will be the average of all {\tt SPC} values.

For this purpose, each code ({\tt SPC, STR, SED}\dots) is allocated a
priority.  The software recognises the code, evaluates the priority, and
calculates an average of the $T_{\rm eff}$ values with highest priority,
ignoring all other values.  The allocation of priority to codes is flexible
and will be different for B stars.  In fact, for B stars only values of
$T_{\rm eff}$ from spectroscopic, narrow-band photometry and the Q method 
should be used.  However, a more pragmatic approach was taken: if 
$T_{\rm eff}$ differs by less than 20\,percent from the value derived from 
the spectral type, it is accepted.  If none of these methods are available, 
then the spectral type serves as an estimate of $T_{\rm eff}$.
  
From time to time, as new sources of $T_{\rm eff}$ become available, the 
master catalogue is updated with the new ``best'' values of $T_{\rm eff}$.  
The priorities used are listed in Table\,\ref{prior}.  The highest priority
is priority 1, the lowest is priority 10.

\begin{table}
\caption{Priority values for effective temperature estimation used in this
work are given in the first two columns, the second column applicable to B
stars only.  The method is listed in the third column while the last column
shows the codes used in the effective temperature catalogue.}
\label{prior}
%\resizebox{8.5cm}{!}{
\begin{tabular}{rrll}
\hline
\multicolumn{1}{c}{p}      & 
\multicolumn{1}{c}{p(B)}   &  
\multicolumn{1}{r}{Method} & 
\multicolumn{1}{l}{Code} \\  
\hline
 1 &  1 &  Spectroscopy, interferometry & SPC,INF     \\
 2 &  2 &  Narrow-band photometry      & BCD,STR,GEN \\
 3 &  3 &  KIC photometry              & KIC         \\
 4 &  4 &  TIC and EPIC photometry     & TIC,EPI     \\
 5 &  5 &  Low-dispersion spectroscopy & LMT         \\
 6 &  6 &  \citep{Anders2022b}         & PAT         \\
 7 &  7 &  SED                         & SED         \\
 8 &  - &  Wide-band photometry        & PHT,IFM     \\
 - &  3 &  Q-method                    & BV0         \\
10 & 10 &  Spectral type               &             \\
\hline           
\end{tabular}    
%}               
\end{table}     

Table\,\ref{teffcat} shows an extract from the effective temperature
catalogue..  The catalogue contains the RA and Dec (J2000), the star name the 
effective temperature and its error (or zero if this is unknown), the 
bibliographic code and the code used in assigning priorities. 

In compiling this catalogue, the {\tt PASTEL} catalogue \citep{Soubiran2016} 
provided a very important starting point. Subsequent entries were made by 
searching the {\tt SIMBAD} database.  About 6\,percent of the effective 
temperatures from \citet{Anders2022b}, which are based on fits to multicolour 
photometry (coded as {\tt PAT} in the table), are inconsistent with the 
spectral type and were rejected.

\begin{table}
\caption{An extract from the effective temperature catalogue.  The full catalogue is
available electronically.}
\label{teffcat}
\resizebox{8.5cm}{!}{
\begin{tabular}{rrlrrll}
\hline
\multicolumn{1}{c}{RA}   & 
\multicolumn{1}{c}{Dec}  &  
\multicolumn{1}{c}{Name}  &  
\multicolumn{1}{r}{$T_{\rm eff}$} & 
\multicolumn{1}{r}{e$T_{\rm eff}$} & 
\multicolumn{1}{c}{Bibcode} & 
\multicolumn{1}{l}{Code} \\  
\hline
   0.3480992 &  19.0760842 & TYC 1181-837-1 & 6066 &     0 & 2022yCat.1354....0A  & PAT \\
   0.3481996 &  30.8280173 & TIC 83957915   & 6138 &     0 & 2018AJ....156..102S  & TIC \\
   0.3481996 &  30.8280173 & TIC 83957915   & 6233 &     0 & 2022yCat.1354....0A  & PAT \\
   0.3486018 &  39.6107104 & HD 224873      & 5147 &    90 & 2009A\&A...504..829G & SPC \\
   0.3486018 &  39.6107104 & HD 224873      & 5386 &    80 & 2011A\&A...530A.138C & SPC \\
   0.3498366 &  34.2818478 & StKM 2-1808    & 4263 &     0 & 2022yCat.1354....0A  & PAT \\
   0.3510299 &   0.2515648 & HD 224877      & 6125 &     0 & 2022yCat.1354....0A  & PAT \\
   0.3510299 &   0.2515648 & HD 224877      & 6156 &   134 & 2020AJ....160..120J  & SPC \\
   0.3519965 & -30.6495347 & TIC 70741976   & 6612 &     0 & 1994MNRAS.268..119B  & LAB \\
   0.3519965 & -30.6495347 & TIC 70741976   & 6727 &     0 & 2018AJ....156..102S  & TIC \\
   0.3519965 & -30.6495347 & TIC 70741976   & 6775 &     0 & 1995A\&A...293...75E & SPC \\
   0.3519965 & -30.6495347 & TIC 70741976   & 7013 &     0 & 2022yCat.1354....0A  & PAT \\
\hline           
\end{tabular}    
}               
\end{table}     

The projected rotational velocity, $v \sin i$, is essential in studies of 
stellar rotation.  For this purpose, the compilation by \citet{Glebocki2005b}
is very important, but rather old.  For this purpose, a literature search for 
values of $v\sin i$ published after this date was initiated.  The updated 
catalogue contains over 75\,000 entries for nearly 52\,000 stars.  Unlike the 
effective temperature catalogue, the adopted value of $v \sin i$ is simply 
taken as the mean value of all measurements for the particular star.
An extract of this catalog is shown in Table\,\ref{vsinicat}.  An entry of
zero for the error in $v\sin i$ means that no error is listed in the
reference.

\begin{table}
\caption{An extract from the projected rotational velocity catalogue.  The full catalogue is
available electronically.}
\label{vsinicat}
\resizebox{8.5cm}{!}{
\begin{tabular}{rrlrrl}
\hline
\multicolumn{1}{c}{RA}   & 
\multicolumn{1}{c}{Dec}  &  
\multicolumn{1}{c}{Name}  &  
\multicolumn{1}{r}{$v\sin i$} & 
\multicolumn{1}{r}{e$v\sin i$} & 
\multicolumn{1}{c}{Bibcode} \\
\hline
 280.2494688 &  43.9151077 & KIC 8073705    &                3.6 &   0.0 & 2018AJ....156..254W  \\
 280.2494688 &  43.9151077 & KIC 8073705    &                3.6 &   1.0 & 2017AJ....154..107P  \\
 280.2510393 &  54.9261528 & TIC 359632996  &                8.0 &   0.0 & 2004A\&A...418..989N \\
 280.2765993 & -26.7405823 & HD 172407      &                9.0 &   0.0 & 2004A\&A...418..989N \\
 280.2906696 &  43.9269491 & KIC 8073767    &               48.4 &  17.0 & 2022A\&A...662A..66X \\
 280.2913298 &  27.9122106 & HD 336659      &               29.2 &   0.0 & 2011ApJ...732...39C  \\
 280.2915445 &  43.9257190 & KIC 8073771    &               55.5 &  15.2 & 2022A\&A...662A..66X \\
 280.3092843 & -41.4399169 & HD 172283      &                9.0 &   0.0 & 2004A\&A...418..989N \\
 280.3192664 & -63.5343778 & HD 171825      &                2.0 &   0.0 & 2004A\&A...418..989N \\
 280.3348041 &   4.5581620 & HD 172675      &                6.0 &   0.0 & 2004A\&A...418..989N \\
 280.3405802 &   7.9036620 & HD 172718      &                5.0 &   0.0 & 2004A\&A...418..989N \\
 280.3408131 &  19.5564334 & HD 349207      &              108.2 &  19.7 & 2022A\&A...662A..66X \\
\hline           
\end{tabular}    
}               
\end{table}     

\section{Effective temperature and luminosity from the spectral type}

Luminosities for all stars in the master catalogue were derived from 
{\it GAIA DR3} parallaxes \citep{Gaia2016,Gaia2022} using the {\em Gaia} 
$G$ magnitude and corresponding bolometric correction from \citet{Chen2019}.  
The interstellar extinction in $G$ is taken from \citet{Anders2022b} which is 
based on a 3D extinction map by \citet{Green2019}.  For the few stars that are 
not listed in the catalogue by \citet{Anders2022b}, the 3D extinction map by 
\citet{Gontcharov2017} was used.  The typical error in $\log L/L_\odot$ is
0.07\,dex which includes the error in the parallax, an error of 0.02\,mag in
$G$, and an error of 0.05\,mag in both the extinction, $A_G$, and the 
bolometric correction. 

Spectral types are mostly from the catalogue of \citet{Skiff2014} supplemented 
by later publications when required.  If more than one spectral type is
available, the one which provides the most information is selected for
inclusion in the master catalogue.  

Given the effective temperature calculated in the manner described above and
the luminosities from the {\em Gaia} parallax, it is possible to derive an
approximate calibration for $T_{\rm eff}$ and $\log L/L_\odot$ as a
function of spectral type and class.  This calibration is shown in 
Table\,\ref{mkclass}.

The relative standard deviation in $T_{\rm eff}$ derived from the spectral type
for luminosity class V is approximately 5\,percent of $T_{\rm eff}$ for stars 
later than B and 10--15\,percent for B stars.  For giants (class III), this is 
about 10--15\,percent for stars later than B, and about 15--20\,percent for B
stars.  The relative standard deviation in luminosity, $\log L/L_\odot$, is
about the same as for $T_{\rm eff}$.

\setlength{\tabcolsep}{10pt}

\begin{table*}
\caption{Calibration of effective temperature and luminosity as a function
of spectral type and class.}
\label{mkclass}
\resizebox{12cm}{!}{
\begin{tabular}{r@{\hskip 10pt}r@{\hskip 1pt}rr@{\hskip 1pt}rr@{\hskip 1pt}
rr@{\hskip 1pt}rr@{\hskip 1pt}r@{\hskip 1pt}}
%\begin{tabular}{rrrrrrrrrrr}
\hline
 &
\multicolumn{2}{c}{V} & \multicolumn{2}{c}{IV} & \multicolumn{2}{c}{III} &
\multicolumn{2}{c}{II} & \multicolumn{2}{c}{I}\\
 & $T_{\rm eff}$ &$\log\tfrac{L}{L_\odot}$
 & $T_{\rm eff}$ &$\log\tfrac{L}{L_\odot}$
 & $T_{\rm eff}$ &$\log\tfrac{L}{L_\odot}$
 & $T_{\rm eff}$ &$\log\tfrac{L}{L_\odot}$
 & $T_{\rm eff}$ &$\log\tfrac{L}{L_\odot}$\\
\hline
 O3 &  37500 &  5.50  &   37500 &  5.50 &   37500 &  5.50 &   36200 &  5.50 &   35000 &  5.50 \\
 O4 &  36600 &  5.34  &   36600 &  5.44 &   36600 &  5.44 &   34600 &  5.44 &   33400 &  5.50 \\
 O5 &  35000 &  5.24  &   35000 &  5.37 &   35000 &  5.50 &   33000 &  5.50 &   31800 &  5.50 \\
 O6 &  34100 &  5.10  &   34100 &  5.25 &   34100 &  5.40 &   31400 &  5.45 &   30300 &  5.50 \\
 O7 &  32800 &  4.92  &   32800 &  5.05 &   32800 &  5.28 &   29800 &  5.30 &   28700 &  5.30 \\
 O8 &  30700 &  4.71  &   30700 &  4.85 &   30700 &  5.01 &   28200 &  5.02 &   27100 &  5.02 \\
 O9 &  28700 &  4.51  &   28700 &  4.76 &   28700 &  4.90 &   26600 &  4.95 &   25600 &  5.00 \\
 B0 &  25900 &  4.24  &   25900 &  4.47 &   25900 &  4.59 &   25000 &  4.62 &   24000 &  4.65 \\
 B1 &  23200 &  3.97  &   23200 &  4.20 &   23200 &  4.46 &   23400 &  4.52 &   22400 &  4.60 \\
 B2 &  22500 &  3.68  &   22500 &  3.88 &   22500 &  4.07 &   21800 &  4.25 &   20800 &  4.40 \\
 B3 &  20800 &  3.48  &   20000 &  3.62 &   19000 &  3.83 &   20200 &  3.91 &   19200 &  4.09 \\
 B4 &  18600 &  3.20  &   18200 &  3.41 &   17900 &  3.63 &   18600 &  3.90 &   17700 &  4.10 \\
 B5 &  17500 &  3.05  &   17000 &  3.18 &   16200 &  3.30 &   17000 &  3.50 &   16100 &  3.90 \\
 B6 &  15400 &  2.74  &   15200 &  2.87 &   15000 &  3.02 &   15400 &  3.40 &   14600 &  3.90 \\
 B7 &  14200 &  2.62  &   14000 &  2.74 &   13700 &  2.83 &   13800 &  3.30 &   13000 &  3.90 \\
 B8 &  12600 &  2.31  &   12600 &  2.49 &   12600 &  2.66 &   12200 &  3.20 &   11400 &  3.88 \\
 B9 &  11300 &  1.99  &   11300 &  2.20 &   11300 &  2.41 &   10600 &  3.00 &    9830 &  3.88 \\
 A0 &   9170 &  1.63  &    9170 &  1.76 &    9170 &  1.92 &    9000 &  2.60 &    8860 &  3.87 \\
 A1 &   8740 &  1.44  &    8740 &  1.50 &    8740 &  1.62 &    8420 &  2.40 &    8100 &  3.86 \\
 A2 &   8500 &  1.42  &    8500 &  1.49 &    8500 &  1.56 &    8290 &  2.30 &    8000 &  3.85 \\
 A3 &   8260 &  1.35  &    8260 &  1.42 &    8260 &  1.49 &    8150 &  2.30 &    7900 &  3.84 \\
 A4 &   8210 &  1.32  &    8210 &  1.39 &    8210 &  1.46 &    8020 &  2.30 &    7800 &  3.83 \\
 A5 &   7950 &  1.25  &    7950 &  1.32 &    7950 &  1.40 &    7890 &  2.30 &    7700 &  3.82 \\
 A6 &   7930 &  1.25  &    7930 &  1.34 &    7930 &  1.44 &    7750 &  2.30 &    7600 &  3.82 \\
 A7 &   7790 &  1.21  &    7790 &  1.28 &    7790 &  1.35 &    7620 &  2.30 &    7500 &  3.81 \\
 A8 &   7630 &  1.19  &    7630 &  1.28 &    7630 &  1.37 &    7490 &  2.30 &    7400 &  3.81 \\
 A9 &   7610 &  1.12  &    7610 &  1.22 &    7610 &  1.31 &    7360 &  2.30 &    7300 &  3.80 \\
 F0 &   7410 &  1.03  &    7410 &  1.16 &    7410 &  1.29 &    7200 &  2.30 &    7200 &  3.80 \\
 F1 &   7210 &  1.03  &    7210 &  1.14 &    7210 &  1.25 &    7090 &  2.30 &    7100 &  3.70 \\
 F2 &   6900 &  0.94  &    6900 &  1.08 &    7030 &  1.23 &    6960 &  2.30 &    6900 &  3.65 \\
 F3 &   6710 &  0.86  &    6710 &  1.03 &    6910 &  1.21 &    6820 &  2.30 &    6800 &  3.60 \\
 F4 &   6830 &  0.87  &    6830 &  1.04 &    6620 &  1.20 &    6690 &  2.32 &    6700 &  3.55 \\
 F5 &   6570 &  0.76  &    6570 &  0.98 &    6510 &  1.21 &    6560 &  2.35 &    6600 &  3.50 \\
 F6 &   6510 &  0.72  &    6510 &  0.92 &    6450 &  1.23 &    6420 &  2.35 &    6500 &  3.45 \\
 F7 &   6370 &  0.61  &    6370 &  0.91 &    6380 &  1.25 &    6290 &  2.36 &    6200 &  3.40 \\
 F8 &   6270 &  0.54  &    6270 &  0.85 &    6280 &  1.27 &    6160 &  2.36 &    6000 &  3.35 \\
 F9 &   6190 &  0.47  &    6190 &  0.79 &    6180 &  1.28 &    6020 &  2.35 &    5900 &  3.35 \\
 G0 &   6040 &  0.43  &    6040 &  0.68 &    6040 &  1.35 &    5890 &  2.37 &    5800 &  3.32 \\
 G1 &   5890 &  0.40  &    5890 &  0.65 &    5890 &  1.49 &    5760 &  2.41 &    5700 &  3.33 \\
 G2 &   5740 &  0.38  &    5740 &  0.65 &    5740 &  1.70 &    5620 &  2.50 &    5600 &  3.30 \\
 G3 &   5600 &  0.34  &    5600 &  0.59 &    5600 &  1.83 &    5490 &  2.55 &    5500 &  3.30 \\
 G4 &   5450 &  0.29  &    5450 &  0.60 &    5450 &  1.95 &    5360 &  2.60 &    5400 &  3.25 \\
 G5 &   5300 &  0.27  &    5300 &  0.68 &    5300 &  2.09 &    5230 &  2.70 &    5300 &  3.25 \\
 G6 &   5150 &  0.24  &    5150 &  0.76 &    5150 &  2.31 &    5090 &  2.75 &    5150 &  3.20 \\
 G7 &   5000 &  0.20  &    5000 &  0.89 &    5000 &  2.51 &    4960 &  2.85 &    5000 &  3.20 \\
 G8 &   4850 &  0.17  &    4850 &  1.08 &    4850 &  2.60 &    4800 &  2.87 &    4850 &  3.15 \\
 G9 &   4700 &  0.14  &    4700 &  1.30 &    4700 &  2.66 &    4690 &  2.90 &    4700 &  3.10 \\
 K0 &   4900 &  0.10  &    4900 &  1.37 &    4114 &  2.72 &    4114 &  2.93 &    4114 &  3.15 \\
 K1 &   4760 & -0.35  &    4760 &  0.89 &    4057 &  2.21 &    4057 &  3.36 &    4057 &  5.00 \\
 K2 &   4620 & -0.40  &    4620 &  0.92 &    4000 &  2.31 &    4000 &  3.39 &    4000 &  5.10 \\
 K3 &   4480 & -0.48  &    4480 &  0.97 &    3900 &  2.40 &    3900 &  3.44 &    3900 &  5.40 \\
 K4 &   4340 & -0.51  &    4340 &  1.01 &    3800 &  2.54 &    3800 &  3.50 &    3800 &  5.50 \\
 K5 &   4200 & -0.58  &    4200 &  1.03 &    3700 &  2.64 &    3700 &  3.56 &    3700 &  5.48 \\
 K6 &   4060 & -0.67  &    4060 &  1.09 &    3717 &  2.67 &    3717 &  3.55 &    3717 &  5.47 \\
 K7 &   3920 & -0.73  &    3920 &  1.15 &    3733 &  2.70 &    3733 &  3.54 &    3733 &  5.42 \\
 K8 &   3780 & -0.81  &    3780 &  1.23 &    3750 &  2.73 &    3750 &  3.57 &    3750 &  4.65 \\
 K9 &   3640 & -0.84  &    3640 &  1.32 &    3725 &  2.78 &    3725 &  3.58 &    3725 &  5.42 \\
 M0 &   3500 & -0.90  &    3500 &  1.42 &    3700 &  2.97 &    3700 &  3.77 &    3700 &  5.50 \\
 M1 &   3333 & -1.21  &    3333 &  1.55 &    3510 &  2.97 &    3510 &  3.77 &    3510 &  5.53 \\
 M2 &   3167 & -1.49  &    3167 &  1.71 &    3320 &  3.08 &    3320 &  3.92 &    3320 &  5.68 \\
 M3 &   3000 & -1.75  &    3000 &  1.89 &    3130 &  3.27 &    3130 &  4.15 &    3130 &  5.87 \\
 M4 &   2833 & -1.98  &    2833 &  2.10 &    2940 &  3.44 &    2940 &  4.36 &    2940 &  6.04 \\
 M5 &   2667 & -2.21  &    2667 &  2.32 &    2750 &  3.71 &    2750 &  4.67 &    2750 &  6.31 \\
 M6 &   2500 & -2.35  &    2500 &  2.61 &    2560 &  3.98 &    2560 &  4.98 &    2560 &  6.62 \\
 M7 &   2333 & -2.43  &    2333 &  2.97 &    2370 &  4.32 &    2370 &  5.40 &    2370 &  6.96 \\
 M8 &   2167 & -2.45  &    2167 &  3.39 &    2180 &  4.71 &    2180 &  5.87 &    2180 &  7.43 \\
 M9 &   2000 & -2.38  &    2000 &  3.90 &    1990 &  5.25 &    1990 &  6.45 &    1990 &  8.01 \\
\hline
\end{tabular}
}
\end{table*}

\begin{figure*}
\centering
\includegraphics[]{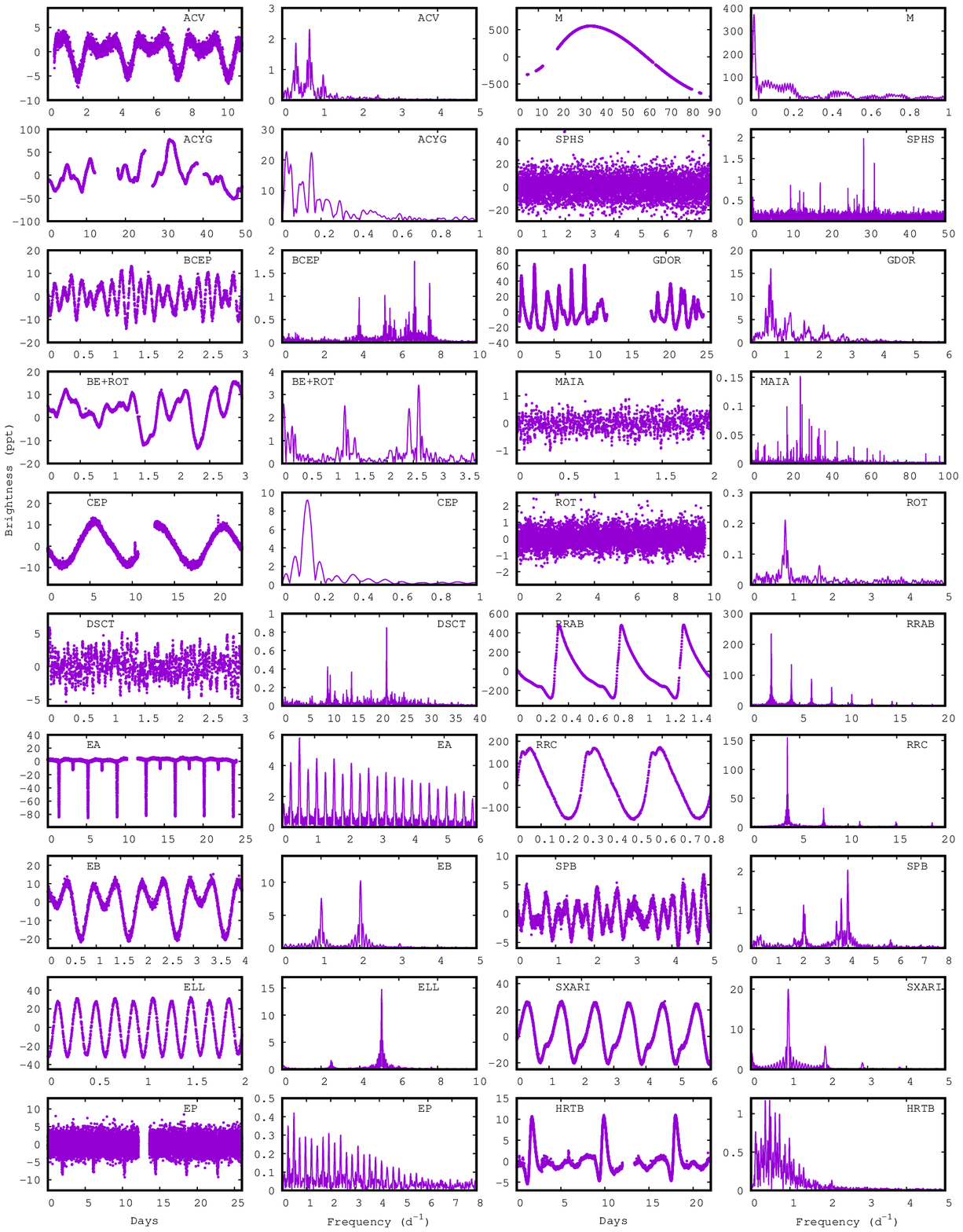}
\caption{Examples of light curves and periodograms for different classes of
variable stars observed by {\em TESS}.  The brightness scale is parts per
thousand.}
\label{curves}
\end{figure*}

\section{The light curve and periodogram}

The {\tt MAST} website provides a bulk download script for all light curves in 
a particular {\em TESS} sector.  The star number in this script is compared 
with the TIC number in the master catalogue.  If there is a match, the
light curve is downloaded as a FITS file.  Typically, about 1000--4000 light
curve files are downloaded per sector.

The next step is to extract the light curve from the FITS file to a  text file.
It is important to be able to generate the full path name of the light curve
from the TIC number. The light curve of TIC\,010510382, for example, is stored 
in file  {\tt T010/010510382/010510382.dat}. The text file contains three 
columns: the time (BJD), the simple aperture (SAP) light curve and the 
pre-search data conditioning (PDC) light curve in parts per thousand.  If the 
star has previously been observed, the latest light curve is simply appended to
the existing light curve.  This procedure has been automated using suitable 
software.  

The next step is calculation of the periodogram and extraction of
significant frequencies, which is usually done by successive prewhitening.
While this technique is perfectly adequate for ground-based light curves, 
problems arise when the signal-to-noise ratio, S/N, is very high, as it is in 
space photometry \citep{Balona2014d}.  In addition, successive prewhitening is 
very slow and impractical for the task at hand.

This problem was solved by mapping the observed time series to a time series 
which is spaced by exactly 2\,min, the spacing in the {\em TESS} observations.  
Where gaps exist in the time series, the data value 
is set to zero.  This allows the periodogram to be calculated using the fast 
Fourier transform (FFT).  For this purpose the number of data points needs to 
be a power of two.  Normally, one would take the power of two just greater 
than the number of points in the time series and pad the excess with zeros.  
However, for greater resolution of the periodogram peaks, the time series was 
extended further by as much as a factor of 4 by zero padding.  In this way
one can obtained well-resolved periodograms of 1000--4000 stars in just a few 
hours.  The periodogram is stored in the same folder as the light curve.

It is well known that the FFT is just another way of representing the data. 
It neither adds nor detracts from the information content. All frequencies
which can be extracted are already visible in the FFT. These can be detected 
simply by noting the peaks in the FFT.  This process is no different from trying 
to extract spectral line wavelengths in a crowded stellar spectrum.  One picks 
the strongest lines, fits suitable curves to each line profile and  removes 
them. Repeating the process one more time is sufficient to extract most of the 
remaining significant spectral wavelengths.  The code which applies the FFT 
technique to the light curve also extracts the frequencies and amplitudes in 
the same way, the window function being used to model each peak.

As each frequency is extracted, the S/N ratio is calculated by measuring the
amplitude relative to a smooth curve representing the periodogram background.  
The extracted information, which is stored in the same folder as the light 
curve and periodogram, contains the frequency, amplitude and their respective 
errors as well as the S/N.  Peaks are deemed significant if S/N $> 4.7$.
After the periodogram for each star has been calculated, variability 
classification may begin.  In this process, the periodogram plays a dominant
role.  In fact, in most cases it is not really possible to perform visual 
classification without the periodogram.  

Before describing the next stage in the classification process, it is first
necessary to discuss all the variability classes that were encountered.

\section{Variability classes}

Variability classification is a subjective process; the distinction between
variability classes is often not clear unless strict guidelines are 
provided.  The informal manner in which these classes were described for
ground-based observations is no longer adequate in the age of space
photometry.  

For example, the $\gamma$~Dor class was originally defined as ``F stars with 
multiple low frequencies''.  This was perfectly adequate because the only F 
stars with multiple frequencies known at that time were the $\delta$~Sct stars.
Since only high frequencies were observed  in $\delta$~Sct stars (in agreement 
with the models), this definition of the $\gamma$~Dor stars presented no problem.  
The definition of ``low frequencies'' was typically taken as frequencies below 
about 5\,d$^{-1}$ (approximately the lowest frequency observed in
ground-based observations of $\delta$~Sct stars), but the exact value was not
part of the definition. In order to drive low frequencies in $\gamma$~Dor 
stars, a different mechanism from that in $\delta$~Sct stars was required 
\citep{Guzik2000}, cementing the distinction between the two classes.

However, the distinction soon became blurred with the discovery of
$\delta$~Sct+$\gamma$~Dor hybrids, i.e. $\delta$~Sct stars with both high
and low frequencies \citep{Handler2002}.  The few hybrids then known suggested 
that these were stars on the boundary between the two classes and could still 
be accommodated by theory.  Later, when {\em Kepler} observations became 
available, it was found that practically all $\delta$~Sct stars contain low 
frequencies \citep{Grigahcene2010}.  

The situation became even less clear when {\em Gaia} parallaxes became 
available.  It was found that $\delta$~Sct stars occupy the same region of the 
H--R diagram as $\gamma$~Dor stars \citep{Balona2018c}.  It seems that 
$\gamma$~Dor stars are not a separate class after all.  Under the circumstances, 
it is difficult to know whether one should perhaps rename the $\gamma$~Dor class 
in a way which unites it with the $\delta$~Sct class.  Until a suitable model 
can be found to explain the low frequencies in $\delta$~Sct stars and how high 
frequencies are damped in $\gamma$~Dor stars, the problem will remain 
unresolved.  

The same situation has recently been found in regard to the relationship
between $\delta$~Sct and roAp stars.  It turns out that many ostensibly
non-peculiar stars pulsate with the high frequencies seen in roAp stars
(above 60\,d$^{-1}$).  Furthermore, the fraction of $\delta$~Sct stars among
Ap stars is no different from that in non-peculiar stars.  Moreover, these 
$\delta$~Sct Ap stars have frequency peaks covering the whole frequency range.
In other words, the roAp stars are just an arbitrary selection of Ap 
$\delta$~Sct stars with no particular significance \citep{Balona2022a}.
 
The significance perhaps lies in a subset of roAp stars in which
equally-spaced multiplets, with spacing equal to the rotation frequency,
are generated due to the effect of the strong magnetic field.  This occurs 
at high frequencies when the pulsational wavelength is comparable to the 
atmospheric scale height.  Perhaps it is more appropriate to 
restrict the name ``roAp'' to these stars.  In any case, the class is probably 
too entrenched and we continue to use it for now to avoid confusion.  
The definition adopted here for roAp is any Ap star with frequencies in excess 
of 60\,d$^{-1}$.  

These high frequencies are also present in non-peculiar $\delta$~Sct stars 
where they were temporarily labeled ``roA'' stars by \citet{Balona2022a}, but 
this name is now dropped. Historically, the upper 
frequency limit for unstable modes in purely radiative models of $\delta$~Sct 
stars is about 50\,d$^{-1}$.  Here we label all $\delta$~Sct stars with 
independent frequencies higher than 50\,d$^{-1}$ as DSCTH (i.e. $\delta$~Sct 
high frequency) and those with independent frequencies higher than 
60\,d$^{-1}$ DSCTU ($\delta$~Sct ultra-high frequency).  An independent
frequency is one which is not a linear combination of two lower frequencies or
multiples of lower frequencies. Whether or not the subclasses DSCTH and DSCTU 
prove useful will have to be decided once we have a theory that can explain 
these pulsations.   

Yet another example of the inadequacy of the current system was the realization
that what appear to be $\gamma$~Dor stars  are to be found at much higher 
effective temperatures than predicted by the models \citep{Balona2016c}.  These
hot $\gamma$~Dor stars are to be found all along the upper main sequence and 
merge with what would be generally described as SPB stars \citep{Balona2020a}.  
In order to avoid this confusion it is necessary to define an effective 
temperature which separates the two groups.  This was taken to be $T_{\rm eff} 
= 10\,000$\,K in \citet{Balona2020a}, a value that is continued in this paper.

Note that many stars are assigned more than one variability class.  For
example, a $\delta$~Sct star which is also suspected to show rotational
peaks is classified as DSCT+ROT.  The criterion used in pulsating stars with
suspected rotational peaks is that the first harmonic must be clearly
present and both peaks must be of reasonably high amplitude.  

The notation SPB/BCEP means that the star may be either classified as an SPB or
BCEP because it falls within the defined $T_{\rm eff}$ range for both stars, 
but the frequency peaks straddle the defined limits for both classes.  In other
words ``+'' means ``and''; ``/'' means ``or''. 

A brief discussion of all variability classes used in this work now follows.
Where there is a possibility of confusion with another class, the adopted
limits on $T_{\rm eff}$ and frequency are given.  This means that what is
classified as a $\gamma$~Dor star may be re-classified as an SPB star, or
vice-versa, if there is a change in the adopted $T_{\rm eff}$.  Pictorial 
examples of what the light curves and periodograms might look like are shown 
in Fig.\,\ref{curves}.

\subsection{Dash}
When a dash is allocated to a variability class it simply means that no
significant variability was detected.  It is also sometimes allocated to a 
star where there seems to be some indistinct, complex low-frequency variations 
which cannot be classified.

\subsection{Pulsating variables}

$\alpha$~Cyg (ACYG) Variables are supergiants with low frequencies
which may be of pulsational origin or perhaps some other cause
\citep{Bowman2019a}.  The light often shows smooth irregular variations, but 
may also be variable on shorter timescales.  In the GCVS the designation is 
limited to B and A supergiants, but here we extend it to F supergiants as well.

$\beta$~Cep (BCEP) are pulsating O8--B6 main sequence stars with multiple periods in the
range 0.1--0.6\,d.  The GCVS also defines a group of short-period
$\beta$~Cep stars (BCEPS), but this class is obsolete.  The long period end
of this definition is more appropriate to SPB stars and in this work BCEP
stars are limited to multiple frequencies in excess of 3\,d$^{-1}$ and
with $T_{\rm eff} > 18000$\,K \citep{Balona2020a}.  The class BCEPH denotes a 
BCEP with one or more frequencies higher than 50\,d$^{-1}$.

The Maia variables (MAIA) are main sequence mid- to late B-type 
stars which pulsate in multiple high frequencies very similar to $\delta$~Sct 
stars \citep{Balona2015c,Balona2016c}.  MAIAH and MAIAU are Maia variables with
frequencies higher than 50 and 60\,d$^{-1}$ respectively. The Maia stars merge 
with the $\beta$~Cep stars at high temperatures and with the $\delta$~Sct stars
at lower temperatures. For this reason, the limit $10000 < T_{\rm eff} < 
18000$\,K was imposed to properly define the MAIA class.  At low frequencies, 
Maia variables merge with the SPB stars.  Therefore it is also necessary to 
impose a frequency limit of 5\,d$^{-1}$.  It has been suggested that Maia stars
are just rapidly rotating SPB stars \citep{Salmon2014}, but this is not 
supported by $v\sin i$ measurements \citep{Balona2020a}. The pulsations may be 
due to high degree modes \citep{Daszynska-Daszkiewicz2017b}.

In the GCVS the slowly pulsating B stars (SPB stars) are labeled as LPB
(long-period pulsating B stars).  In this paper we use the name SPB.  These 
are multiperiodic mid- and late-B stars with frequencies less than about
5\,d$^{-1}$.  However, {\em TESS} has shown that these variables are present 
among all main-sequence B stars, including the hottest B stars 
\citep{Balona2020a}.  They can be confused with the $\beta$~Cep variables, 
many of which have frequencies as low as 3\,d$^{-1}$.  In order to 
avoid this confusion, the designation SPB is used for stars with 
$T_{\rm eff} > 18000$\,K with frequencies less than 3\,d$^{-1}$.  For cooler B 
stars ($10000 < T_{\rm eff} < 18000$\,K), the frequency limit is extended to 5\,d$^{-1}$ to avoid confusion with 
Maia variables \citep{Balona2020a}.  The SPB and $\gamma$~Dor stars seem to form a 
continuous sequence across all B, A and F main sequence stars for which no
explanation exists at this time.  For this reason, the SPB class is
restricted to stars hotter than 10\,000\,K.

The $\gamma$~Dor stars (GDOR), $\delta$~Sct (DSCT) and roAp stars have already
been discussed.  To avoid confusion with $\delta$~Sct and SPB star, the 
frequencies in GDOR stars must be lower than 5\,d$^{-1}$ and 
$T_{\rm eff} < 10\,000$\,K.  The $\delta$~Sct stars must have $T_{\rm eff} < 
10\,000$\,K to avoid confusion with Maia variables. 

SX~Phe (SXPHE) stars resemble the $\delta$~Sct variables but belong to spherical component
or old disk galactic population.  They are mainly present in globular
clusters.

Solar-type oscillators (SOLR) are generally F or K stars with very low 
amplitude pulsations similar to the 5-min solar oscillations.  Large numbers 
of these stars were detected by {\em Kepler}, but are much more difficult to 
identify with {\em TESS} owing to the larger noise level.  The SOLR stars 
listed in this paper are all {\em Kepler} stars.

Cepheids (CEP) are F and G supergiants with radial pulsations.  Later spectral 
types have longer periods.  The GCVS has three other subclass, CEP(B), DCEP 
and DCEPS, but these are not used here.

The RR~Lyraes (RR, RRAB, RRC) are radially pulsating Population II giants.  
They are present in large numbers in globular clusters.  The general 
designation is RR, but most of the stars observed by {\em Kepler} and 
{\em TESS} are of the RRAB class with asymmetric light curves with a steep 
ascending branch.  A few RRC stars, which have symmetrical almost sinusoidal 
light curves, have also been observed by {\em TESS}.

The W~Vir stars (CW) are Population II pulsating variable).  The subclasses 
CWA and CWB are W~Vir stars with periods respectively longer and shorter than
8\,d.

The RV Tau (RV) stars are radially pulsating supergiants with F/G spectral 
types at maximum and K/M at minimum with alternating primary and secondary 
minima. The minima vary in depth so that the primary minimum may become the 
secondary minimum and vice-versa.  The period between two primary minima is 
30--150\,d.  The subclass RVA does not vary in mean magnitude, while RVB does 
vary in mean magnitude.

The Mira (M) variables are long-period variable giants with characteristic 
late-type emission spectra (Me, Ce, Se) and semi-regular pulsations with 
periods 80--1000\,d. 

Semiregular variables (SR) are giants or supergiants of intermediate and late 
spectral types showing noticeable light periodicity (from 20 to over 2000\,d), 
but with irregular interruptions.  The GCVS divides these into subclasses
SRA (persistent periodicity) and SRB (poorly defined periodicity).  The SRD
class are F, G or K supergiants with 30--1100\,d periods.

In the GCVS the V361~Hya stars are labeled RPHS (rapidly pulsating hot subdwarfs), 
but there is no class for the slowly pulsating V1093~Her stars. The label
SPHS is used here.  The separation between the two classes is taken to be about 
300\,d$^{-1}$ in frequency, which means that it is not possible to detect
pure RPHS stars with {\em TESS}.  Those that are labeled RPHS are taken from
the literature.   Originally, there seemed to be a distinction in 
effective temperature between RPHS and SPHS stars, the SPHS being cooler.
This distinction seems to have disappeared with further observations 
\citep{Krzesinski2022}.  Most pulsating B subdwarfs appear to be hybrids 
(SPHS+RPHS).

PVTEL variables are helium supergiant Bp stars with weak hydrogen lines and 
enhanced lines of He and C.  They pulsate with periods of approximately
0.1--1\,d.  The BLAP (blue large-amplitude pulsator) are a proposed new class 
of degenerate star \citep{Pietrukowicz2013}. At this stage little is known 
about them.
   
ZZ Ceti variables (ZZ) are nonradially pulsating white dwarfs with periods in the
range 0.5--25\,min.  Flares are sometimes observed which are usually
attributed to a cool companion.  The ZZA are of DA spectral type, while ZZB
are DB stars.  The ZZO class was introduced in the GCVS for ZZ Cet variables of
the DO spectral type showing HeII and and CIV absorption lines in their spectra.
This also includes the GW~Vir class.

\subsection{Rotational variables}

In the GCVS, the BY~Dra (BY) variables are emission-line dwarfs of dKe-dMe spectral 
type showing quasi-periodic light changes with periods from a fraction of a day 
to 120\,d. The light variability is caused by spots.  Many such star also
flare.  The FK Com stars (FKCOM) are rapidly-rotating G or K giants with 
non-uniform surface brightness.  The light variation is due to rotation.

In the GCVS the $\alpha^2$~CVn (ACV) stars are defined by a strictly periodic 
variation in B8p--A7p stars. They exhibit magnetic field and brightness changes
(periods of 0.5--160 days or more).  These peculiar main-sequence stars 
display strong lines of Si, Sr, Cr and rare earths. The light modulation is 
attributed to surface patches with differing abundances.  In this work, the ACV 
designation is applied to any Ap/Fp star with $T_{\rm eff} < 10000$\,K which
shows strict periodicity compatible with the expected rotation frequency.

In the GCVS, SX~Ari (SXARI) variables are main sequence peculiar B stars with 
variable intensity He\,I and Si\,III lines and strong magnetic fields. The 
light period is the same as the rotational period.  These are the 
high-temperature analogs of the ACV variables.   In this paper, SX~Ari stars
are peculiar B stars with $T_{\rm eff} > 10000$\,K.  We distinguish 
between the SX~Ari stars and the He-weak and He-rich variables in that SXARI 
stars must have anomalous abundances of Si or Hg/Mn.  This is a subject which 
requires a deeper study.  The He-variable stars are usually classified as ROT 
if they have a strict light period consistent with rotation.

The label MRP describes stars which emit radio pulses, two per rotation period 
\citep{Das2022}.  They appear to be ACV or SXARI stars.

Even a brief examination of {\em Kepler} or {\em TESS} photometry will show
that a significant fraction of A and B stars have an isolated low frequency,
low amplitude peak at less than 4\,d$^{-1}$, consistent with what one might 
expect for the rotational frequency.  The harmonic of the peak is often 
present.   Many studies have shown that these are indeed to be identified as 
the rotational frequency (see \citealt{Balona2022b} and references therein). 
The GCVS does not have a designation for hot stars with rotational
modulation. For this reason, the ROT class was introduced.  A subclass, the 
ROTD stars, is reserved for  stars which show a sharp rotational peak at a 
slightly higher frequency than a broad peak which is suspected to be due to 
inertial modes (the ``hump and spike'' stars \citealt{Balona2014b,Saio2018a}). 
It is often not possible to distinguish between rotational modulation and
low-level eclipses and for this reason the ROT class is restricted to
frequency peaks with amplitudes less than 10\,ppt.  However, most of the ROT
variables have amplitudes less than 0.5\,ppt \citep{Balona2022b}.

\subsection{Eclipsing variables}

The general GCVS classification for eclipsing stars is E.  However,
more specific subclasses have been introduced.  Note that not
much attention has been payed to specific subclasses of eclipsing variables
in this work.  Generally, any star in which eclipses are suspected are
classified as EA or EB.

The EA class, or Algol ($\beta$~Per)-class eclipsing systems, are binaries 
with spherical or slightly ellipsoidal components. This class is used when it 
is possible to specify the moments of the beginning and end of the eclipses. 
Between eclipses the light remains almost constant or varies insignificantly 
because of reflection effects.

The EB ($\beta$~Lyr) eclipsing systems have ellipsoidal components and light 
curves for which it is impossible to specify the exact times of onset and end 
of eclipses. The depth of secondary minimum is usually considerably smaller 
than that of the primary minimum.

AM Her variables (AM) are close binary systems consisting of a dK or dM star
and a compact object with a strong magnetic field. 

DM are detached main sequence binary systems in which the inner Roche lobes
are not filled.

The HRTB (heartbeat stars, \citealt{Welsh2011}) are variable binary star 
systems in eccentric orbits with pulsations driven by tidal forces.

The ellipsoidal variables (ELL) are close binary systems with ellipsoidal 
components.  The combined brightness varies with a period equal to the orbital 
period, but there are no eclipses.

In the GCVS the designation EP is given to stars with eclipses due to planets.
In this work we use EP for any eclipse-like variation with an amplitude less
than about 10\,ppt.  The periodogram typically contains several low-amplitude
harmonics which could either be due to low-level eclipses or rotation.  The
distinction is made mainly on the number of visible harmonics.  If only one
or two harmonics are seen, then the star is classed as ROT, otherwise as EP.
It is quite possible that many stars classified as EP are actually
rotational variables or even planetary systems.  An analysis of these stars for
planetary systems would be interesting as very few A and B stars are known
to host planets. 

W UMa class eclipsing variables (EW) have periods shorter than 1\,d and the
primary and secondary minima are equal or almost equal in depth.  In the
GCVS this designation is used only for F or later type stars.

GS eclipsing systems are giants or supergiants where one of the components
may be a main sequence star.

The RS CVn stars (RS) are eruptive variables in a binary system showing Ca II H
and K emission.  The light period is close to the orbital period and they are
X-ray sources.

\begin{table*}
\caption{A list of {\em Kepler, K2} and {\em TESS} stars classified by the
author. The full catalogue is available electronically.  The column labeled p 
is the priority number for the adopted effective temperature. The rotation 
period, $P_{\rm rot}$ is given where applicable.  The column marked pec is a 
code for the stellar peculiarity.}
\label{mastercat}
\resizebox{18cm}{!}{
\begin{tabular}{rrllrrrrrrl}
\hline
\multicolumn{1}{c}{RA}   & 
\multicolumn{1}{c}{Dec}  &  
\multicolumn{1}{c}{Name} & 
\multicolumn{1}{l}{Var}  & 
\multicolumn{1}{r}{$T_{\rm eff}$} & 
\multicolumn{1}{r}{$\log\tfrac{L}{L_\odot}$} & 
\multicolumn{1}{r}{p}    &
\multicolumn{1}{r}{$v\sin i$} & 
\multicolumn{1}{r}{$P_{\rm rot}$} & 
\multicolumn{1}{r}{pec} & 
\multicolumn{1}{l}{Sp. Type} \\  
\hline
 279.8504662 & -47.1368375 & TIC 339907982  & EB      & 31000 &  1.62 &  4&        &         &  9 & sdB,sdB,sdO/BsdB  \\ 
 279.8654723 &  25.1250158 & TIC 316904504  & -       &  7970 &  1.20 &  4&        &         &    & A0                \\
 279.8681614 &  45.3642605 & TIC 383751023  & -       &  7819 &  1.09 &  1&        &         &  1 & kA2hA6mF4         \\
 279.8695086 &  43.5142894 & KIC 7797592    & ROT     &  6905 &  0.83 &  3&        &   1.326 &    & [F5V]             \\
 279.8759807 & -21.9655564 & EPIC 216652481 & ROT/SPB & 12600 &  3.39 & 10&  174.0 &   1.786 &  6 & B8IIIe            \\
 279.8769148 & -48.5085760 & TIC 304451865  & DSCT    &  7289 &  1.14 &  4&        &         &    & A8/9V             \\
 279.8783231 &  18.3319300 & TIC 345978356  & -       &  5971 &  0.11 &  6&        &         &    & G0                \\
 279.8834254 & -55.3068028 & TIC 120202126  & ROT     &  6449 &  0.57 &  4&        &   4.651 &    & F3V               \\
 279.8867664 & -72.9482134 & TIC 344167219  & -       &  6559 &  0.46 &  4&        &         &    & F3V               \\
 279.8876674 &  40.9350562 & TIC 157696928  & SPB     & 10423 &  2.51 &  1&  106.0 &         &    & B8:Vn + A0III     \\
\hline           
\end{tabular}    
}               
\end{table*}

\subsection{Eruptive and irregular variables}

The designation FLARE is attached to any object where evidence of a flaring
event is present in the light curve.

The Be stars (BE, BE+ROT) are eruptive  B stars in which Balmer line emission 
is seen,  though the emission might disappear and re-appear at different times.
Recent observations from {\em TESS} have shown that about 75\,percent of Be
stars vary with short quasi-periods in the range 0.5--2.0\,d \citep{Balona2020b}.
It is suggested that the eruptions and subsequent rotational modulation is a
result of localized surface activity (flaring) which ejects material.  This
material is channelled to diametrically opposed regions where the geometric
and magnetic equators intersect.  The light and line-profile variation is
due to obscuration by these trapped clouds which eventually disperse into
the circumstellar disc \citep{Balona2021d, Balona2022b}.
    
Both the periodogram and the light curve are very distinctive.  The periodogram
normally shows a low-frequency group (suspected to be a result of
circumstellar material) as well as two other groups.  One group can be
identified at about the expected rotational frequency, while the other group 
is at about twice the rotational frequency.  This is a consequence of
obscuration by the diametrically located trapped clouds.  The amount of
trapped material varies with time, sometimes resulting in the disappearance
of the highest frequency group.  What is clear is that the peaks are broad and 
vary in phase and amplitude with a timescale of weeks to months 
\citep{Balona2020b, Balona2021d}. The light curve often shows distinctive
eclipse-like dips, quite different from the light curves of SPB and
$\gamma$~Dor stars which vary in the same period range, but with strict
coherent frequencies.

In this work, Be stars which show the three groups are labeled as BE+ROT 
while those which only show long-period variations are classed as BE.  The ROT 
designation refers to the possible non-coherent rotational light variation of 
the two higher-frequency groups as described.  The distinctive periodogram and 
light curve is sometimes seen in ostensibly non-Be stars.  These may be 
incipient Be stars which are designated BEV or BEV+ROT.  

The $\gamma$~Cas (GCAS) stars are Be stars with X-ray emission.
The $\sigma^2$~Ori variables (SORI) are Be stars in which the light
variation is due to circumstellar matter trapped by a strong tilted dipole
magnetic field.
Most Wolf-Rayet (WR) stars show irregular light variations, but periodicity is
present in a few stars (WR+BCEP or WR+ROT).

UV are variables of the UV Ceti class, which are KVe--MVe stars sometimes 
displaying flare activity with amplitudes from several tenths of a magnitude 
up to several magnitudes.

The S~Doradus (SDOR) are eruptive, high luminosity Bp--Fp stars showing irregular
(sometimes cyclic) light changes with amplitudes 1--7\,mag in V.  As a rule,
these stars are associated with diffuse nebulae and surrounded by expanding
envelopes.

U Geminorum-class variables (UG), quite often called dwarf novae, are close 
binary systems consisting of a dwarf or subgiant K-M star that fills the volume
of its inner Roche lobe and a white dwarf surrounded by an accretion disk. 
Orbital periods are in the range 0.05-0.5\,d.  These are divided into 
subclasses UGSU (SU~UMa systems) and UGSS (SS~Cyg systems) and UGZ (Z~Cam
systems).

Stars of variability class I are an homogeneous set of poorly studied 
irregular variables.  This can be divided into class IA (early-type, O--A) and 
IB (late type).  The IN type are irregular, eruptive variables connected with 
bright or dark diffuse nebulae or observed in the regions of these nebulae.  
These again are divided into early type (INA) and late type (INB).  The INS and
ISA classes show rapid light variations.  INT or IT are Orion variables of the 
T~Tauri type. NL are poorly studied eruptive nova-like variables.  PN stars
are the nuclei of planetary nebulae.  The LB class describes late-type stars 
with slow irregular variations.

The NB, NC and NL classes are novae with fast (NA), slow (NB) and very slow
(NC) light variations.  The NL class are poorly studied nova-like
variables.         

XP denotes an X-ray pulsar source.  The subclass XNG describes a nova-like
transient system consisting an early-type giant or supergiant with a compact
companion.

\section{Classification}

Once the {\em TESS} stars have been downloaded and the periodograms
calculated, the next step is to display the light curves and periodograms 
of unclassified stars for visual classification.  It is imperative that 
these curves be displayed with the utmost ease to reduce the time required 
for classification.  For this purpose, software was developed which 
enabled the display of these curves by simply clicking on a list of TIC 
star numbers.  

The periodogram is the most important source of information for classifying 
purposes.  The process requires two passes.  In the first pass, the full 
frequency range of the periodogram is displayed (0--360\,d$^{-1}$) so that 
high frequencies can be detected.  In this pass, many $\delta$~Sct stars 
and eclipsing binaries are normally seen.  In the second pass, the displayed 
frequency range is narrowed to 0--10\,d$^{-1}$.  This pass proceeds more 
slowly as other classes of variable are identified.  In every new {\em TESS} release there are around 1000--2000 
previously unclassified stars.  The process of classification is usually 
accomplished in 2--4\,days.

Among the stars in the master catalogue observed by {\em Kepler, K2} and
{\em TESS}, only about 2\,percent appear in the GCVS. The variability class 
derived by visual inspection is nearly always in good agreement with the GCVS 
class, but corrections were sometimes made.

An extract of the catalog, which is complete up to {\em TESS} sector 55, is
shown in Table\,\ref{mastercat}.  In this table, stars without spectral type
have been assigned spectral types in accordance with the measured $T_{\rm
eff}$ and $\log L/L_\odot$.  These spectral types are enclosed in square
brackets.  The stellar peculiarity has been assigned a code in accordance
with the following scheme:
 1: Am (CP1); 2: Ap (CP2);  3: Hg/Mn (CP3);
 4: He-weak (CP4);  5: He-rich (CP5); 6: Classical Be;
 7: Herbig Be; 8: $\lambda$~Boo; 9: sdOB.

\begin{figure}
\centering
\includegraphics[]{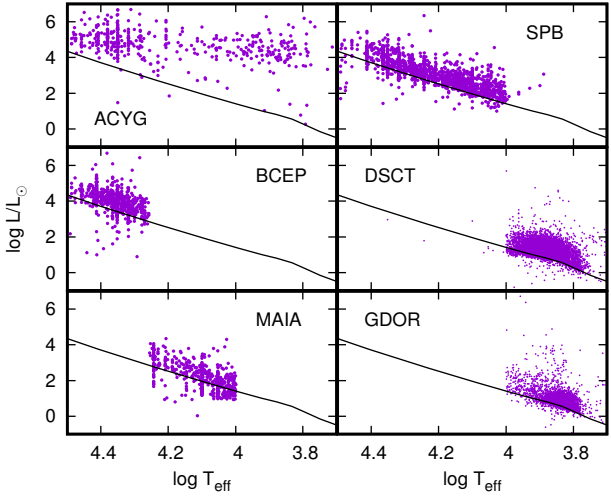}
\caption{Location of some pulsating stars in the H--R diagram. The solid
line is the zero-age main sequence from solar composition models by
\citet{Bertelli2009}. }
\label{hrdiag}
\end{figure}

There are 20\,744 stars observed by {\em Kepler}, 5\,877 stars observed by 
{\em K2} and 94\,186 {\em TESS} stars (total 120\,807 stars).
The location of several types of pulsating star in the H-R diagram is shown
in Fig.\,\ref{hrdiag}.

\begin{table}
\caption{Number of stars in the most common classes and sub-classes.}
\label{stats}
%\resizebox{18cm}{!}{
\begin{tabular}{lrlr}
\hline
\multicolumn{1}{l}{Class} & 
\multicolumn{1}{c}{N}     & 
\multicolumn{1}{l}{Class} & 
\multicolumn{1}{c}{N}     \\ 
\hline
TOTAL           &    120807 &  FLARE (all)    &      1146 \\
                &           &                 &           \\                            
Non-variable (-)&     45064 & Be  (all)       &       723 \\ 
                &           & BE+ROT          &       589 \\ 
Rotational (all)&     42983 & SORI            &        15 \\ 
ROT             &      1302 & BEV             &        49 \\
ROTD            &       289 & GCAS            &         3 \\
BY              &         8 &                 &           \\
FKCOM           &         2 & BCEP (all)      &       700 \\
ACV             &       996 &                 &           \\
SXARI           &       675 & Maia (all)      &       518 \\
                &           & MAIA            &       475 \\
$\delta$~Sct (all) &     13755 & MAIAH           &        28 \\
DSCT            &     10164 & MAIAU           &        15 \\
DSCTH           &      1468 &                 &           \\
DSCTU           &      2001 & ACYG (all)      &       426 \\
ROAP            &       121 &                 &           \\
                &           & Tidal (all)     &       459 \\
GDOR (all)      &      6128 & ELL             &       374 \\
                &           & HRTB            &        85 \\
Eclipsing (all):&      5928 &                 &           \\
EA              &      2722 & SPHS            &       178 \\
EB              &      1996 &                 &           \\
EP              &       970 & CEP             &       132 \\ 
EW              &       216 & CW              &        16 \\
                &           &                 &           \\ 
SOLR (all)      &      2415 & RRAB            &       198 \\
SPB (all)       &      1286 & RRC             &        52 \\
\hline           
\end{tabular}    
%}               
\end{table}

\section{Conclusions}

Over the last decade, the author has been visually classifying the light
variations of stars observed by {\em Kepler, K2} and {\em TESS}.  The
classification system has been kept as close as possible to that of the {\em
General Catalogue of Variable Stars}, but a few modifications were
inevitable.  The results are presented as a table of over 120\,000 stars.
These data could be used as a learning sample for an artificial intelligence 
code which can then be applied to all {\em TESS} light curves.  
Table\,\ref{stats} summarizes the number of stars in each class and
subclass for the most common classes.

The informal and rather vague definitions of variability classes in the {\em
General Catalogue of Variable Stars} is no longer appropriate to the large
number of stars observed from space photometry.  In this work we have
attempted to define the classes according to limits of effective temperature
and frequency, where necessary.  There is a pressing need for a more
comprehensive theoretical understanding of stellar pulsation along the main
sequence.  It is quite clear that preconceived assumptions regarding the
atmospheres in upper main sequence stars need to be relaxed.  Until the
models are able to replicate the locations of the stars in the H--R diagram
and their pulsational frequencies, no progress can be made in understanding
the different classes of pulsational variables that have been observed.

In order to support this project, the author has also been collecting
estimates of effective temperature from the literature.  This is presented
as a table of over 900\,000 individual measurements for nearly 635\,000
stars with literature references and a code which indicates the method used.  A 
catalogue of projected rotational velocities containing over 75\,000 individual
measurements for nearly 52\,000 stars with bibliographic reference code is 
also presented.

The assigned variability class represents the author's opinion and it is
possible that others would assign a different class for many stars.  There
is no ``correct'' variability class.  There are, no doubt, errors in these
three catalogues which have escaped the author's attention and there is no
claim of completeness in any of these catalogues.  They are presented
in the hope that they will prove useful to the community.

\section*{Acknowledgments}

LAB wishes to thank the National Research Foundation of South Africa for 
financial support. 

This paper includes data collected by the {\it TESS} mission. Funding for the 
{\it TESS} mission is provided by the NASA Explorer Program. Funding for the 
{\it TESS} Asteroseismic Science Operations Centre is provided by the Danish 
National Research Foundation (Grant agreement no.: DNRF106), ESA PRODEX
(PEA 4000119301) and Stellar Astrophysics Centre (SAC) at Aarhus University. 
We thank the {\it TESS} and TASC/TASOC teams for their support of the present
work.

This work has made use of data from the European Space Agency (ESA) mission
{\it Gaia} (\url{https://www.cosmos.esa.int/gaia}), processed by the {\it Gaia}
Data Processing and Analysis Consortium (DPAC,
\url{https://www.cosmos.esa.int/web/gaia/dpac/consortium}). Funding for the DPAC
has been provided by national institutions, in particular the institutions
participating in the {\it Gaia} Multilateral Agreement.
This research has made use of the SIMBAD database, operated at CDS, 
Strasbourg, France.

The data presented in this paper were obtained from the Mikulski Archive for 
Space Telescopes (MAST).  STScI is operated by the Association of Universities
for Research in Astronomy, Inc., under NASA contract NAS5-2655.

This research has made use of the SIMBAD database, operated at CDS, Strasbourg, 
France.

\section*{Data availability}

All data are incorporated into the article and its online supplementary
material. The data are available at:\\
\\
{\tt https://sites.google.com/view/tessvariables/home}

\bibliographystyle{mn2e}
\bibliography{var}

\begin{thebibliography}{39}
\expandafter\ifx\csname natexlab\endcsname\relax\def\natexlab#1{#1}\fi

\bibitem[{{Anders} {et~al.}(2022){Anders}, {Khalatyan}, {Queiroz}, {Chiappini},
  {Ard{\`e}vol}, {Casamiquela}, {Figueras}, {Jim{\'e}nez-Arranz}, {Jordi},
  {Mongui{\'o}}, {Romero-G{\'o}mez}, {Altamirano}, {Antoja}, {Assaad},
  {Cantat-Gaudin}, {Castro-Ginard}, {Enke}, {Girardi}, {Guiglion}, {Khan},
  {Luri}, {Miglio}, {Minchev}, {Ramos}, {Santiago}, \&
  {Steinmetz}}]{Anders2022b}
{Anders} F., {Khalatyan} A., {Queiroz} A.~B.~A., {et~al.}, 2022, \aap, 658, A91

\bibitem[{{Audenaert} {et~al.}(2021){Audenaert}, {Kuszlewicz}, {Handberg},
  {Tkachenko}, {Armstrong}, {Hon}, {Kgoadi}, {Lund}, {Bell}, {Bugnet},
  {Bowman}, {Johnston}, {Garc{\'\i}a}, {Stello}, {Moln{\'a}r}, {Plachy},
  {Buzasi}, {Aerts}, \& {T'DA Collaboration}}]{Audenaert2021}
{Audenaert} J., {Kuszlewicz} J.~S., {Handberg} R., {et~al.}, 2021, \aj, 162,
  209

\bibitem[{{Balona}(2014{\natexlab{a}})}]{Balona2014d}
{Balona} L.~A., 2014{\natexlab{a}}, \mnras, 439, 3453

\bibitem[{{Balona}(2014{\natexlab{b}})}]{Balona2014b}
---, 2014{\natexlab{b}}, \mnras, 441, 3543

\bibitem[{{Balona}(2018)}]{Balona2018c}
---, 2018, \mnras, 479, 183

\bibitem[{{Balona}(2022{\natexlab{a}})}]{Balona2022a}
---, 2022{\natexlab{a}}, \mnras, 510, 5743

\bibitem[{{Balona}(2022{\natexlab{b}})}]{Balona2022b}
---, 2022{\natexlab{b}}, \mnras, 516, 3641

\bibitem[{{Balona} {et~al.}(2015){Balona}, {Baran},
  {Daszy{\'n}ska-Daszkiewicz}, \& {De Cat}}]{Balona2015c}
{Balona} L.~A., {Baran} A.~S., {Daszy{\'n}ska-Daszkiewicz} J., {De Cat} P.,
  2015, \mnras, 451, 1445

\bibitem[{{Balona} {et~al.}(2016){Balona}, {Engelbrecht}, {Joshi}, {Joshi},
  {Sharma}, {Semenko}, {Pandey}, {Chakradhari}, {Mkrtichian}, {Hema}, \&
  {Nemec}}]{Balona2016c}
{Balona} L.~A., {Engelbrecht} C.~A., {Joshi} Y.~C., {et~al.}, 2016, \mnras,
  460, 1318

\bibitem[{{Balona} \& {Ozuyar}(2020{\natexlab{a}})}]{Balona2020a}
{Balona} L.~A., {Ozuyar} D., 2020{\natexlab{a}}, \mnras, 493, 5871

\bibitem[{{Balona} \& {Ozuyar}(2020{\natexlab{b}})}]{Balona2020b}
---, 2020{\natexlab{b}}, \mnras, 493, 2528

\bibitem[{{Balona} \& {Ozuyar}(2021)}]{Balona2021d}
---, 2021, \apj, 921, 5

\bibitem[{{Bertelli} {et~al.}(2009){Bertelli}, {Nasi}, {Girardi}, \&
  {Marigo}}]{Bertelli2009}
{Bertelli} G., {Nasi} E., {Girardi} L., {Marigo} P., 2009, \aap, 508, 355

\bibitem[{{Blomme} {et~al.}(2010){Blomme}, {Debosscher}, {De Ridder}, {Aerts},
  {Gilliland}, {Christensen-Dalsgaard}, {Kjeldsen}, {Brown}, {Borucki}, {Koch},
  {Jenkins}, {Kurtz}, {Stello}, {Stevens}, {Suran}, \& {Derekas}}]{Blomme2010}
{Blomme} J., {Debosscher} J., {De Ridder} J., {et~al.}, 2010, \apjl, 713, L204

\bibitem[{{Blomme} {et~al.}(2011){Blomme}, {Sarro}, {O'Donovan}, {Debosscher},
  {Brown}, {Lopez}, {Dubath}, {Rimoldini}, {Charbonneau}, {Dunham},
  {Mandushev}, {Ciardi}, {De Ridder}, \& {Aerts}}]{Blomme2011}
{Blomme} J., {Sarro} L.~M., {O'Donovan} F.~T., {et~al.}, 2011, \mnras, 418, 96

\bibitem[{{Bowman} {et~al.}(2019){Bowman}, {Burssens}, {Pedersen}, {Johnston},
  {Aerts}, {Buysschaert}, {Michielsen}, {Tkachenko}, {Rogers}, {Edelmann},
  {Ratnasingam}, {Sim{\'o}n-D{\'\i}az}, {Castro}, {Moravveji}, {Pope}, {White},
  \& {De Cat}}]{Bowman2019a}
{Bowman} D.~M., {Burssens} S., {Pedersen} M.~G., {et~al.}, 2019, Nature
  Astronomy, 3, 760

\bibitem[{{Brown} {et~al.}(2011){Brown}, {Latham}, {Everett}, \&
  {Esquerdo}}]{Brown2011a}
{Brown} T.~M., {Latham} D.~W., {Everett} M.~E., {Esquerdo} G.~A., 2011, \aj,
  142, 112

\bibitem[{{Chen} {et~al.}(2019){Chen}, {Girardi}, {Fu}, {Bressan}, {Aringer},
  {Dal Tio}, {Pastorelli}, {Marigo}, {Costa}, \& {Zhang}}]{Chen2019}
{Chen} Y., {Girardi} L., {Fu} X., {et~al.}, 2019, \aap, 632, A105

\bibitem[{{Das} {et~al.}(2022){Das}, {Chandra}, {Shultz}, {Wade}, {Sikora},
  {Kochukhov}, {Neiner}, {Oksala}, \& {Alecian}}]{Das2022}
{Das} B., {Chandra} P., {Shultz} M.~E., {et~al.}, 2022, \apj, 925, 125

\bibitem[{{Daszy{\'n}ska-Daszkiewicz}
  {et~al.}(2017){Daszy{\'n}ska-Daszkiewicz}, {Walczak}, \&
  {Pamyatnykh}}]{Daszynska-Daszkiewicz2017b}
{Daszy{\'n}ska-Daszkiewicz} J., {Walczak} P., {Pamyatnykh} A., 2017, in
  European Physical Journal Web of Conferences, Vol. 160, European Physical
  Journal Web of Conferences, p. 03013

\bibitem[{{Gaia Collaboration} {et~al.}(2022){Gaia Collaboration}, {Gaia
  Collaboration}, {De Ridder}, {Ripepi}, {Aerts}, {Palaversa}, {Eyer}, {Holl},
  {Audard}, {Rimoldini}, {Brown}, {Vallenari}, {Prusti}, {de Bruijne},
  {Arenou}, {Babusiaux}, {Biermann}, {Creevey}, {Ducourant}, {Evans}, {Guerra},
  {Hutton}, {Jordi}, {Klioner}, {Lammers}, {Lindegren}, {Luri}, {Mignard},
  {Panem}, {Pourbaix}, {Randich}, {Sartoretti}, {Soubiran}, {Tanga}, {Walton},
  {Bailer-Jones}, {Bastian}, {Drimmel}, {Jansen}, {Katz}, {Lattanzi}, {van
  Leeuwen}, {Bakker}, {Cacciari}, {Casta{\~n}eda}, {De Angeli}, {Fabricius},
  {Fouesneau}, {Fr{\'e}mat}, {Galluccio}, {Guerrier}, {Heiter}, {Masana},
  {Messineo}, {Mowlavi}, {Nicolas}, {Nienartowicz}, {Pailler}, {Panuzzo},
  {Riclet}, {Roux}, {Seabroke}, {Sordo}, {Th{\'e}venin}, {Gracia-Abril},
  {Portell}, {Teyssier}, {Altmann}, {Andrae}, {Bellas-Velidis}, {Benson},
  {Berthier}, {Blomme}, {Burgess}, {Busonero}, {Busso}, {C{\'a}novas}, {Carry},
  {Cellino}, {Cheek}, {Clementini}, {Damerdji}, {Davidson}, {de Teodoro},
  {Nu{\~n}ez Campos}, {Delchambre}, {Dell'Oro}, {Esquej},
  {Fern{\'a}ndez-Hern{\'a}ndez}, {Fraile}, {Garabato}, {Garc{\'\i}a-Lario},
  {Gosset}, {Haigron}, {Halbwachs}, {Hambly}, {Harrison}, {Hern{\'a}ndez},
  {Hestroffer}, {Hilger}, {Hodgkin}, {Jan{\ss}en}, {Jevardat de Fombelle},
  {Jordan}, {Krone-Martins}, {Lanzafame}, {L{\"o}ffler}, {Marchal}, {Marrese},
  {Moitinho}, {Muinonen}, {Osborne}, {Pancino}, {Pauwels}, {Recio-Blanco},
  {Reyl{\'e}}, {Riello}, {Roegiers}, {Rybizki}, {Sarro}, {Siopis}, {Smith},
  {Sozzetti}, {Utrilla}, {van Leeuwen}, {Abbas}, {{\'A}brah{\'a}m}, {Abreu
  Aramburu}, {Aguado}, {Ajaj}, {Aldea-Montero}, {Altavilla}, {{\'A}lvarez},
  {Alves}, {Anders}, {Anderson}, {Anglada Varela}, {Antoja}, {Baines}, {Baker},
  {Balaguer-N{\'u}{\~n}ez}, {Balbinot}, {Balog}, {Barache}, {Barbato},
  {Barros}, {Barstow}, {Bartolom{\'e}}, {Bassilana}, {Bauchet}, {Becciani},
  {Bellazzini}, {Berihuete}, {Bernet}, {Bertone}, {Bianchi}, {Binnenfeld},
  {Blanco-Cuaresma}, {Boch}, {Bombrun}, {Bossini}, {Bouquillon}, {Bragaglia},
  {Bramante}, {Breedt}, {Bressan}, {Brouillet}, {Brugaletta}, {Bucciarelli},
  {Burlacu}, {Butkevich}, {Buzzi}, {Caffau}, {Cancelliere}, {Cantat-Gaudin},
  {Carballo}, {Carlucci}, {Carnerero}, {Carrasco}, {Casamiquela}, {Castellani},
  {Castro-Ginard}, {Chaoul}, {Charlot}, {Chemin}, {Chiaramida}, {Chiavassa},
  {Chornay}, {Comoretto}, {Contursi}, {Cooper}, {Cornez}, {Cowell}, {Crifo},
  {Cropper}, {Crosta}, {Crowley}, {Dafonte}, {Dapergolas}, {David}, {de
  Laverny}, {De Luise}, {De March}, {de Souza}, {de Torres}, {del Peloso}, {del
  Pozo}, {Delbo}, {Delgado}, {Delisle}, {Demouchy}, {Dharmawardena}, {Diakite},
  {Diener}, {Distefano}, {Dolding}, {Enke}, {Fabre}, {Fabrizio}, {Faigler},
  {Fedorets}, {Fernique}, {Figueras}, {Fournier}, {Fouron}, {Fragkoudi}, {Gai},
  {Garcia-Gutierrez}, {Garcia-Reinaldos}, {Garc{\'\i}a-Torres}, {Garofalo},
  {Gavel}, {Gavras}, {Gerlach}, {Geyer}, {Giacobbe}, {Gilmore}, {Girona},
  {Giuffrida}, {Gomel}, {Gomez}, {Gonz{\'a}lez-N{\'u}{\~n}ez},
  {Gonz{\'a}lez-Santamar{\'\i}a}, {Gonz{\'a}lez-Vidal}, {Granvik}, {Guillout},
  {Guiraud}, {Guti{\'e}rrez-S{\'a}nchez}, {Guy}, {Hatzidimitriou}, {Hauser},
  {Haywood}, {Helmer}, {Helmi}, {Sarmiento}, {Hidalgo}, {H{\l}adczuk}, {Hobbs},
  {Holland}, {Huckle}, {Jardine}, {Jasniewicz}, {Jean-Antoine Piccolo},
  {Jim{\'e}nez-Arranz}, {Juaristi Campillo}, {Julbe}, {Karbevska}, {Kervella},
  {Khanna}, {Kordopatis}, {Korn}, {K{\'o}sp{\'a}l}, {Kostrzewa-Rutkowska},
  {Kruszy{\'n}ska}, {Kun}, {Laizeau}, {Lambert}, {Lanza}, {Lasne}, {Le
  Campion}, {Lebreton}, {Lebzelter}, {Leccia}, {Leclerc}, {Lecoeur-Taibi},
  {Liao}, {Licata}, {Lindstr{\o}m}, {Lister}, {Livanou}, {Lobel}, {Lorca},
  {Loup}, {Madrero Pardo}, {Magdaleno Romeo}, {Managau}, {Mann}, {Manteiga},
  {Marchant}, {Marconi}, {Marcos}, {Marcos Santos}, {Mar{\'\i}n Pina},
  {Marinoni}, {Marocco}, {Marshall}, {Polo}, {Mart{\'\i}n-Fleitas}, {Marton},
  {Mary}, {Masip}, {Massari}, {Mastrobuono-Battisti}, {Mazeh}, {McMillan},
  {Messina}, {Michalik}, {Millar}, {Mints}, {Molina}, {Molinaro}, {Moln{\'a}r},
  {Monari}, {Mongui{\'o}}, {Montegriffo}, {Montero}, {Mor}, {Mora},
  {Morbidelli}, {Morel}, {Morris}, {Muraveva}, {Murphy}, {Musella}, {Nagy},
  {Noval}, {Oca{\~n}a}, {Ogden}, {Ordenovic}, {Osinde}, {Pagani}, {Pagano},
  {Palicio}, {Pallas-Quintela}, {Panahi}, {Payne-Wardenaar}, {Pe{\~n}alosa
  Esteller}, {Penttil{\"a}}, {Pichon}, {Piersimoni}, {Pineau}, {Plachy},
  {Plum}, {Poggio}, {Pr{\v{s}}a}, {Pulone}, {Racero}, {Ragaini}, {Rainer},
  {Raiteri}, {Ramos}, {Ramos-Lerate}, {Re Fiorentin}, {Regibo}, {Richards},
  {Rios Diaz}, {Riva}, {Rix}, {Rixon}, {Robichon}, {Robin}, {Robin}, {Roelens},
  {Rogues}, {Rohrbasser}, {Romero-G{\'o}mez}, {Rowell}, {Royer}, {Ruz Mieres},
  {Rybicki}, {Sadowski}, {S{\'a}ez N{\'u}{\~n}ez}, {Sagrist{\`a} Sell{\'e}s},
  {Sahlmann}, {Salguero}, {Samaras}, {Sanchez Gimenez}, {Sanna},
  {Santove{\~n}a}, {Sarasso}, {Schultheis}, {Sciacca}, {Segol}, {Segovia},
  {S{\'e}gransan}, {Semeux}, {Shahaf}, {Siddiqui}, {Siebert}, {Siltala},
  {Silvelo}, {Slezak}, {Slezak}, {Smart}, {Snaith}, {Solano}, {Solitro},
  {Souami}, {Souchay}, {Spagna}, {Spina}, {Spoto}, {Steele},
  {Steidelm{\"u}ller}, {Stephenson}, {S{\"u}veges}, {Surdej}, {Szabados},
  {Szegedi-Elek}, {Taris}, {Taylor}, {Teixeira}, {Tolomei}, {Tonello}, {Torra},
  {Torra}, {Torralba Elipe}, {Trabucchi}, {Tsounis}, {Turon}, {Ulla}, {Unger},
  {Vaillant}, {van Dillen}, {van Reeven}, {Vanel}, {Vecchiato}, {Viala},
  {Vicente}, {Voutsinas}, {Weiler}, {Wevers}, {Wyrzykowski}, {Yoldas}, {Yvard},
  {Zhao}, {Zorec}, {Zucker}, \& {Zwitter}}]{Gaia2022}
{Gaia Collaboration}, {Gaia Collaboration}, {De Ridder} J., {et~al.}, 2022,
  arXiv e-prints, arXiv:2206.06075

\bibitem[{{Gaia Collaboration} {et~al.}(2016){Gaia Collaboration}, {Prusti},
  {de Bruijne}, {Brown}, {Vallenari}, {Babusiaux}, {Bailer-Jones}, {Bastian},
  {Biermann}, {Evans}, \& et~al.}]{Gaia2016}
{Gaia Collaboration}, {Prusti} T., {de Bruijne} J.~H.~J., {et~al.}, 2016, \aap,
  595, A1

\bibitem[{{Glebocki} \& {Gnacinski}(2005)}]{Glebocki2005b}
{Glebocki} R., {Gnacinski} P., 2005, VizieR Online Data Catalog, 3244

\bibitem[{{Gontcharov}(2017)}]{Gontcharov2017}
{Gontcharov} G.~A., 2017, Astronomy Letters, 43, 472

\bibitem[{{Green} {et~al.}(2019){Green}, {Schlafly}, {Zucker}, {Speagle}, \&
  {Finkbeiner}}]{Green2019}
{Green} G.~M., {Schlafly} E., {Zucker} C., {Speagle} J.~S., {Finkbeiner} D.,
  2019, \apj, 887, 93

\bibitem[{{Grigahc{\`e}ne} {et~al.}(2010){Grigahc{\`e}ne}, {Antoci}, {Balona},
  {Catanzaro}, {Daszy{\'n}ska-Daszkiewicz}, {Guzik}, {Handler}, {Houdek},
  {Kurtz}, {Marconi}, {Monteiro}, {Moya}, {Ripepi}, {Su{\'a}rez},
  {Uytterhoeven}, {Borucki}, {Brown}, {Christensen-Dalsgaard}, {Gilliland},
  {Jenkins}, {Kjeldsen}, {Koch}, {Bernabei}, {Bradley}, {Breger}, {Di
  Criscienzo}, {Dupret}, {Garc{\'{\i}}a}, {Garc{\'{\i}}a Hern{\'a}ndez},
  {Jackiewicz}, {Kaiser}, {Lehmann}, {Mart{\'{\i}}n-Ruiz}, {Mathias},
  {Molenda-{\.Z}akowicz}, {Nemec}, {Nuspl}, {Papar{\'o}}, {Roth}, {Szab{\'o}},
  {Suran}, \& {Ventura}}]{Grigahcene2010}
{Grigahc{\`e}ne} A., {Antoci} V., {Balona} L., {et~al.}, 2010, \apjl, 713, L192

\bibitem[{{Guzik} {et~al.}(2000){Guzik}, {Kaye}, {Bradley}, {Cox}, \&
  {Neuforge}}]{Guzik2000}
{Guzik} J.~A., {Kaye} A.~B., {Bradley} P.~A., {Cox} A.~N., {Neuforge} C., 2000,
  \apjl, 542, L57

\bibitem[{{Handler} {et~al.}(2002){Handler}, {Balona}, {Shobbrook}, {Koen},
  {Bruch}, {Romero-Colmenero}, {Pamyatnykh}, {Willems}, {Eyer}, {James}, \&
  {Maas}}]{Handler2002}
{Handler} G., {Balona} L.~A., {Shobbrook} R.~R., {et~al.}, 2002, \mnras, 333,
  262

\bibitem[{{Krzesinski} \& {Balona}(2022)}]{Krzesinski2022}
{Krzesinski} J., {Balona} L.~A., 2022, \aap, 663, A45

\bibitem[{{Pietrukowicz} {et~al.}(2013){Pietrukowicz}, {Dziembowski},
  {Mr{\'o}z}, {Soszy{\'n}ski}, {Udalski}, {Poleski}, {Szyma{\'n}ski}, {Kubiak},
  {Pietrzy{\'n}ski}, {Wyrzykowski}, {Ulaczyk}, {Koz{\l}owski}, \&
  {Skowron}}]{Pietrukowicz2013}
{Pietrukowicz} P., {Dziembowski} W.~A., {Mr{\'o}z} P., {et~al.}, 2013, \actaa,
  63, 379

\bibitem[{{Pojmanski} {et~al.}(2005){Pojmanski}, {Pilecki}, \&
  {Szczygiel}}]{Pojmanski2005}
{Pojmanski} G., {Pilecki} B., {Szczygiel} D., 2005, \actaa, 55, 275

\bibitem[{{Saio} {et~al.}(2018){Saio}, {Kurtz}, {Murphy}, {Antoci}, \&
  {Lee}}]{Saio2018a}
{Saio} H., {Kurtz} D.~W., {Murphy} S.~J., {Antoci} V.~L., {Lee} U., 2018,
  \mnras, 474, 2774

\bibitem[{{Salmon} {et~al.}(2014){Salmon}, {Montalb{\'a}n}, {Reese}, {Dupret},
  \& {Eggenberger}}]{Salmon2014}
{Salmon} S.~J.~A.~J., {Montalb{\'a}n} J., {Reese} D.~R., {Dupret} M.-A.,
  {Eggenberger} P., 2014, \aap, 569, A18

\bibitem[{{Samus} {et~al.}(2017){Samus}, {Kazarovets}, {Durlevich}, {Kireeva},
  \& {Pastukhova}}]{Samus2017}
{Samus} N.~N., {Kazarovets} E.~V., {Durlevich} O.~V., {Kireeva} N.~N.,
  {Pastukhova} E.~N., 2017, Astronomy Reports, 61, 80

\bibitem[{{Skiff}(2014)}]{Skiff2014}
{Skiff} B.~A., 2014, VizieR Online Data Catalog, 1, 2023

\bibitem[{{Soubiran} {et~al.}(2016){Soubiran}, {Le Campion}, {Brouillet}, \&
  {Chemin}}]{Soubiran2016}
{Soubiran} C., {Le Campion} J.-F., {Brouillet} N., {Chemin} L., 2016, \aap,
  591, A118

\bibitem[{{Stassun} {et~al.}(2018){Stassun}, {Oelkers}, {Pepper}, {Paegert},
  {De Lee}, {Torres}, {Latham}, {Charpinet}, {Dressing}, {Huber}, {Kane},
  {L{\'e}pine}, {Mann}, {Muirhead}, {Rojas-Ayala}, {Silvotti}, {Fleming},
  {Levine}, \& {Plavchan}}]{Stassun2018}
{Stassun} K.~G., {Oelkers} R.~J., {Pepper} J., {et~al.}, 2018, \aj, 156, 102

\bibitem[{{Welsh} {et~al.}(2011){Welsh}, {Orosz}, {Aerts}, {Brown},
  {Brugamyer}, {Cochran}, {Gilliland}, {Guzik}, {Kurtz}, {Latham}, {Marcy},
  {Quinn}, {Zima}, {Allen}, {Batalha}, {Bryson}, {Buchhave}, {Caldwell},
  {Gautier}, {Howell}, {Kinemuchi}, {Ibrahim}, {Isaacson}, {Jenkins}, {Prsa},
  {Still}, {Street}, {Wohler}, {Koch}, \& {Borucki}}]{Welsh2011}
{Welsh} W.~F., {Orosz} J.~A., {Aerts} C., {et~al.}, 2011, \apjs, 197, 4

\bibitem[{{Wenger} {et~al.}(2000){Wenger}, {Ochsenbein}, {Egret}, {Dubois},
  {Bonnarel}, {Borde}, {Genova}, {Jasniewicz}, {Lalo{\"e}}, {Lesteven}, \&
  {Monier}}]{Wenger2000}
{Wenger} M., {Ochsenbein} F., {Egret} D., {et~al.}, 2000, \aaps, 143, 9

\end{thebibliography}

\label{lastpage}

\newpage

\end{document}